\documentclass[11pt, twoside, table]{predoc2}

\usepackage[utf8]{inputenc}
\usepackage[T1]{fontenc}
\usepackage{newtxtext} 
\usepackage[scaled=.95]{cabin} 
\usepackage[varqu,varl]{inconsolata} 
\usepackage{amsmath}
\usepackage{tabularx}
\usepackage{xltabular}
\usepackage[varg]{newtxmath}
\usepackage{changepage}
\usepackage{enumitem}
\usepackage[colorlinks=true,
    linkcolor=red,
    filecolor=black,
    citecolor=black,      
    urlcolor=black]{hyperref}       
\usepackage{url}            
\usepackage{graphicx}
\usepackage{float}
\usepackage{booktabs}       
\usepackage{amsfonts}       
\usepackage{nicefrac}       
\usepackage{microtype}      
\usepackage{xcolor}         

\newcommand{\vectheta}{\mathrm{\boldsymbol{\theta}}}
\newcommand{\thetaspace}{\Theta}

\title{Bridge Sampling Diagnostics}

\author[1]{Giorgio Micaletto\thanks{The majority of this work was completed while the author was working at Aalto University.}}
\affil[1]{Bocconi University, Italy}
\author[2]{Aki Vehtari}
\affil[2]{ELLIS Institute Finland; Department of Computer Science, Aalto University, Finland}

\makeatletter
\def\runauthor{Micaletto and Vehtari}
\def\runtitle{Bridge Sampling Diagnostics}
\makeatother

\begin{document}

\maketitle

\begin{abstract}
\noindent 
In Bayesian statistics, the marginal likelihood is used for model selection and averaging, yet it is often challenging to compute accurately for complex models. Approaches such as bridge sampling, while effective, suffer from high variance when the proposal distribution overlaps poorly with the target posterior.
To quantify this variance, we present a closed-form Monte Carlo standard error (MCSE) estimator for bridge sampling, extending classical variance approximations with a multi-chain effective-sample-size correction for autocorrelated MCMC draws and an exact log-scale variance. We show that the MCSE estimate itself is structurally capped at $\approx 1.05$, so values near this cap signal saturation rather than precision, and our calibration experiments show that the MCSE can be trusted when it is below 0.3.
Furthermore, we introduce a hybrid score-matching proposal that regularizes the sample covariance using the local posterior geometry, significantly improving the stability of the estimator, and we demonstrate the efficacy of these methods using increasingly difficult simulated posteriors and real posteriors from the \texttt{posteriordb} database.
\end{abstract}

\noindent%

\thispagestyle{empty}

\section{Introduction}
\label{sec:introduction}

Marginal likelihoods (model evidences) are often used for Bayesian model comparison (e.g. Bayes factors), model averaging, and calibrating prior and model components. At the same time, it is well understood that evidence-based workflows can be fragile in practice: marginal likelihood estimates can be extremely sensitive to prior specification and to numerical error, with different estimators' behavior varying significantly across models and data. We do not advocate marginal likelihood as a universal model selection tool (see, e.g., \citealp{friel_wyse_2012,fourment_dubious_2019} for broader reviews and cautions). Our goal is instead to reduce wasted computation and failure-by-silence. We provide practical diagnostics for the \emph{Monte Carlo reliability} of evidence estimates computed from a finite number of posterior draws, together with an improved proposal distribution for when the default is not reliable.

Let $\mathcal{M}$ denote a Bayesian model with parameter vector $\vectheta \in \thetaspace\subseteq\mathbb R^{p}$ (where $p \coloneqq \dim(\vectheta)$) and observed data vector $\mathbf{y}=(y_1,\dots,y_n)^\top\in\mathbb{R}^n$.
\begin{equation*}
p(\mathbf{y}\mid \mathcal{M})
= \int_{\thetaspace} p(\mathbf{y}\mid \vectheta,\mathcal{M})\,p(\vectheta\mid\mathcal{M})\, d\vectheta,
\end{equation*}
which is generally intractable for realistic models and thus requires approximation. Among sampling-based methods, bridge sampling has a long history starting from the acceptance-ratio ideas of \citet{bennett_efficient_1976} and the general formulation of \citet{meng_simulating_1996}, with connections to importance sampling and path sampling emphasized by \citet{gelman_meng_1998}. In modern Bayesian workflows bridge sampling has become a practical default, in part due to robust implementations and tutorials (e.g., \citealp{gronau_tutorial_2017,gronau_bridgesampling_2020}).

Despite its practical appeal, bridge sampling is not ``automatic'': a single run at a typical posterior sample size does not always yield a numerically reliable estimate. The estimator is a ratio of empirical averages, and its stability is governed by two things: the overlap between the posterior and the proposal distribution, and the behavior of rare, extreme terms. In high-dimensional or strongly non-Gaussian posteriors, the distribution of the summands can exhibit severe pre-asymptotic behavior, so that a small number of atypical draws may dominate the estimate long before the asymptotic $\sqrt{S}$ regime becomes a useful approximation. Re-running full posterior inference to assess Monte Carlo variability is often computationally expensive, and practitioners therefore greatly benefit from diagnostics that can be computed from existing posterior draws and likelihood evaluations.

We develop a diagnostic and methodological workflow for bridge sampling that aims to answer a simple practical question: \emph{given the posterior draws already available, can we trust the reported marginal likelihood estimate}? Concretely, we make three contributions:
\begin{enumerate}[label=(\roman*)]
    \item We present a Monte Carlo standard error (MCSE) estimator for the bridge sampling estimate and its logarithm. Its relative-variance core is the classical result of \citet{meng_simulating_1996} in the practical form of \citet{fruhwirth-schnatter_estimating_2004}. New here are the multi-chain effective-sample-size autocorrelation adjustment \citep{Vehtari+etal:2021:Rhat}, the propagation to the log scale, and the large-scale empirical calibration of the resulting MCSE.
    \item We show that the MCSE estimate has a structural upper bound of $\approx 1.05$, which it cannot exceed however badly the estimator fails. Values near this saturation cap therefore signal pre-asymptotic failure rather than genuine precision, and we validate the practical trust threshold $\operatorname{MCSE} < 0.3$.
    \item We introduce a hybrid score-matching proposal distribution that utilizes affine-invariant Riemannian geometry to combine the empirical sample covariance with gradient-based local curvature, mitigating pre-asymptotic tail mismatch in moderate-to-high dimensions.
\end{enumerate}
We demonstrate the behavior of these methods using increasingly difficult simulated posteriors, and across a broad suite of real posteriors from the \texttt{posteriordb} database.

\section{Bridge Sampling}
\label{sec:bridgesampling}

Write the unnormalized posterior density as
\[
\widetilde{\pi}(\vectheta)
\coloneqq
p(\mathbf{y}\mid \vectheta,\mathcal{M})\,p(\vectheta\mid \mathcal{M}),
\qquad
\pi(\vectheta)
\;=\;
\frac{\widetilde{\pi}(\vectheta)}{Z},
\qquad
Z \;\coloneqq\; p(\mathbf{y}\mid \mathcal{M})
\;=\; \int_{\thetaspace} \widetilde{\pi}(\vectheta)\, d\vectheta.
\]
Bridge sampling estimates the normalizing constant $Z$ using two ingredients. The first is a \emph{normalized} proposal density $g(\vectheta)$, which we must be able to both sample from and evaluate pointwise, and whose support overlaps that of the posterior $\pi(\vectheta)$. The second is a bridge function $h(\vectheta)$, also denoted $\omega(\vectheta)$ in parts of the literature. Formally, $h(\vectheta)$ is a user-chosen auxiliary function defined on the common support of $\pi$ and $g$ such that
\[
0 < \left| \int_{\thetaspace} h(\vectheta) \pi(\vectheta) g(\vectheta) \, d\vectheta \right| < \infty.
\]
Provided this condition holds, $h(\vectheta)$ acts as an algebraic device satisfying the exact identity
\begin{equation}
\label{eq:expectation-bridgesampling}
Z
\;=\;
\frac{\mathbb{E}_{g}\!\left[\widetilde{\pi}(\vectheta)\,h(\vectheta)\right]}
     {\mathbb{E}_{\pi}\!\left[g(\vectheta)\,h(\vectheta)\right]},
\end{equation}
where 
\[
\mathbb{E}_g[f(\vectheta)] \coloneqq \int_{\thetaspace} f(\vectheta)\, g(\vectheta)\, d\vectheta
\]
and 
\[
\mathbb{E}_\pi[f(\vectheta)] \coloneqq \int_{\thetaspace} f(\vectheta)\, \pi(\vectheta)\, d\vectheta.
\]
Hence, $h$ \emph{defines} the resulting Monte Carlo estimator and strongly affects its variance through how it reweights regions where $\pi$ and $g$ overlap, with many familiar estimators arising as special cases of bridge sampling via particular choices of $h$ (e.g., importance sampling and reciprocal-importance/harmonic-mean–type estimators).

Using the identity \eqref{eq:expectation-bridgesampling}, we obtain the empirical estimator
\begin{equation}
\label{eq:bridgesampling}
\widehat{Z}
\;=\;
\widehat{p}(\mathbf{y}\mid \mathcal{M})
\;=\;
\frac{\frac{1}{S_2}\sum_{i=1}^{S_2}
      \widetilde{\pi}(\widetilde{\vectheta}_i)\,h(\widetilde{\vectheta}_i)}
     {\frac{1}{S_1}\sum_{j=1}^{S_1}
      h(\vectheta_j^*)\,g(\vectheta_j^*)},
\end{equation}
where \(\vectheta_j^\ast\sim\pi(\vectheta)\) are \(S_1\) posterior draws reserved for evaluation and \(\widetilde{\vectheta}_i\sim g(\vectheta)\) are \(S_2\) independent proposal draws. The total available MCMC draws are split into a training subset, which is used to fit the proposal \(g(\vectheta)\), and the remaining \(S_1\) draws, which constitute the evaluation subset used in Equation \eqref{eq:iterative_bridgesampling}. The evaluation draws are autocorrelated within the Markov chain, but independent of the \(S_2\) proposal draws generated from the fitted \(g(\vectheta)\).

\begin{figure}[tp]
  \centering
  \includegraphics[width=0.7\textwidth]{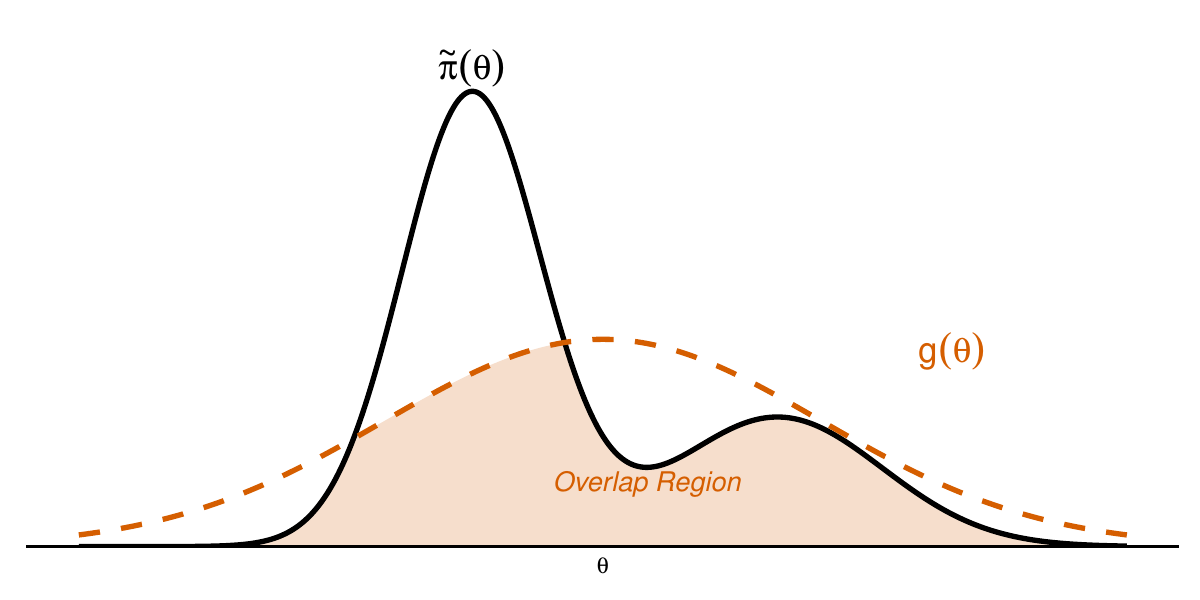}
  \caption{Intuition of the bridge sampling estimator. The estimation of the marginal likelihood relies on the ratio of the unnormalized posterior $\widetilde{\pi}(\vectheta)$ (solid line) and the proposal density $g(\vectheta)$ (dashed line), and its stability depends on the \emph{overlap} (shaded region) between the two distributions.}
  \label{fig:bridge_schematic}
\end{figure}

The practical performance of bridge sampling is governed by the overlap between $\pi(\vectheta)$ and $g(\vectheta)$, illustrated in Figure \ref{fig:bridge_schematic}. In the standard workflow, $g(\vectheta)$ is a multivariate normal distribution in the unconstrained parameter space, with mean and covariance estimated from posterior draws \citep{gronau_tutorial_2017,gronau_bridgesampling_2020}.
Because proposal fitting from posterior draws can introduce bias if the same draws are reused, implementations often split posterior draws into subsets (one subset to fit $g$, one subset to estimate $Z$), trading reduced bias for increased Monte Carlo variability \citep{overstall_forster_2010,wong_splitting_2020}. For a comprehensive tutorial and step-by-step implementation details, we refer the reader to the tutorial by \citet{gronau_tutorial_2017}.

Strongly non-Gaussian or multimodal posteriors overlap poorly with normal proposals. A complementary line of work addresses this by transforming (warping) the target to match a tractable reference distribution more closely, as in warp bridge sampling \citep{meng_schilling_2002, gronau_computing_2020} and in Warp-U--style transformations for multimodal targets \citep{wang_warpU_2022}.
We take a different route: rather than transforming the target, we improve the proposal covariance directly (Section \ref{sec:hybrid-score-matching}). We include Warp-III bridge sampling as a comparison in Section \ref{sec:results}. We also provide diagnostics that indicate when the default workflow is already reliable, and when it is worth investing in more posterior draws, a better proposal, or an alternative evidence estimator.

A widely used (near-optimal) choice of $h(\vectheta)$ is the one proposed by \citet{meng_simulating_1996},
\begin{equation*}
h(\vectheta) \propto
\frac{1}{
s_1\widetilde{\pi}(\vectheta)
+
s_2\,Z\,g(\vectheta)},
\end{equation*}
where
\[
s_1 \coloneqq \frac{S_1}{S_1 + S_2}
\quad\text{and}\quad
s_2 \coloneqq \frac{S_2}{S_1 + S_2}.
\]
More generally, the asymptotically optimal bridge function under independence is proportional to the reciprocal of a mixture of the two unnormalized densities \citep{bennett_efficient_1976,meng_simulating_1996}. This optimal form depends on the unknown $Z$, and is therefore implemented in practice via a fixed-point iteration.
Defining
\[
l_{1,j} \coloneqq \frac{\widetilde{\pi}(\vectheta^*_j)}{g(\vectheta^*_j)},
\qquad
l_{2,i} \coloneqq \frac{\widetilde{\pi}(\widetilde{\vectheta}_i)}{g(\widetilde{\vectheta}_i)},
\]
the standard update takes the form
\begin{equation}
\label{eq:iterative_bridgesampling}
 \widehat Z^{(t+1)}=
 \frac{\dfrac{1}{S_2}\sum_{i=1}^{S_2}\dfrac{l_{2,i}}{s_1l_{2,i}+s_2\widehat Z^{(t)}}}
      {\dfrac{1}{S_1}\sum_{j=1}^{S_1}\dfrac{1}{s_1l_{1,j}+s_2\widehat Z^{(t)}}},
 \qquad \text{with}\;\widehat Z^{(0)}=1,
\end{equation}
iterating until the update is smaller than a tolerance (or until a maximum number of iterations is reached). 

Bridge sampling is typically less sensitive than plain importance sampling to modest tail mismatch because the update in \eqref{eq:iterative_bridgesampling} involves bounded ratios (see, e.g., \citealp{gronau_tutorial_2017}). However, boundedness alone does not guarantee stable finite-sample behavior: when overlap is poor, the effective range of the summands can be so wide that rare extreme realizations dominate at practical sample sizes $S_1$ and $S_2$. The remainder develops MCSE estimation and complementary diagnostics aimed at detecting exactly these pre-asymptotic failure modes without requiring repeated posterior inference.

\section{Monte Carlo Standard Error}
\label{sec:mcse}

As the first step to assess the precision of the bridge sampling estimate, we compute the Monte Carlo standard error (MCSE).
The estimator \(\widehat{Z}\) given in Equation~\eqref{eq:bridgesampling} is a ratio of two sample means (at the fixed point of the iteration): \( \widehat{Z} = \bar{N}/\bar{D}\), with
\begin{equation*}
\begin{aligned}
    \bar{N} &= \frac{1}{S_2}\sum_{i=1}^{S_2}N_i, \quad \text{where}\quad N_i \,\coloneqq\, \widetilde{\pi}(\widetilde{\vectheta}_i) h(\widetilde{\vectheta}_i), \\
    \bar{D} &= \frac{1}{S_1}\sum_{j=1}^{S_1}D_j, \quad \text{where}\quad D_j \,\coloneqq\, h(\vectheta^*_j)\, g(\vectheta^*_j).
\end{aligned}
\end{equation*}
To compute the variance of \(\widehat{Z}\), we use the delta method for a ratio of means:
\begin{equation*}
    \operatorname{Var}\left( \widehat{Z}\right) \approx \left( \frac{\bar{N}}{\bar{D}} \right)^2 \left( \frac{\operatorname{Var}(\bar{N})}{\bar{N}^2} + \frac{\operatorname{Var}(\bar{D})}{\bar{D}^2} - 2\, \frac{\operatorname{Cov}(\bar{N}, \bar{D})}{\bar{N}\, \bar{D}} \right).
\end{equation*}
The three terms simplify as follows. Because the proposal distribution \(g(\vectheta)\) is fitted on a separate split of the posterior draws, the evaluation draws \(\vectheta_j^\ast\) and the generated proposal draws \(\widetilde{\vectheta}_i\) are conditionally independent, so \(\operatorname{Cov}(\bar{N}, \bar{D})=0\). The numerator and the denominator do share the bridge function \(h\), which depends on \(\widehat{Z}\) and hence on both sets of draws. At the fixed point this coupling is of second order. The optimal bridge function minimizes the asymptotic variance, so replacing the unknown \(Z\) in \(h\) by an estimate with \(O(S^{-1/2})\) error perturbs the variance only by \(O(S^{-1})\). The iterated estimator therefore attains the same first-order variance as the infeasible estimator that uses the true \(Z\) \citep[Theorem 2]{meng_simulating_1996}. Consistent with this, we find no detectable numerator--denominator dependence in our experiments. The proposal draws are i.i.d., so \(\operatorname{Var}(\bar{N}) = \operatorname{Var}(N_i)/S_2\). The evaluation draws are MCMC draws, so the terms \(D_j\) may exhibit significant autocorrelation or non-mixing chains; we account for this by replacing the raw sample size \(S_1\) with the effective sample size \(S_{1,\mathrm{eff}}\) \citep{Vehtari+etal:2021:Rhat}, giving \(\operatorname{Var}(\bar{D}) \approx \operatorname{Var}(D_j)/S_{1,\mathrm{eff}}\).

Substituting back, we get
\begin{equation}
\label{eq:var-approx-no-cov}
    \operatorname{Var}\left( \widehat{Z} \right) \approx \left( \frac{\bar{N}}{\bar{D}} \right)^2 \left( \frac{\operatorname{Var}(N_i)}{S_2\, \bar{N}^2} + \frac{\operatorname{Var}(D_j)}{S_{1,\mathrm{eff}}\, \bar{D}^2} \right).
\end{equation}
To turn Equation \eqref{eq:var-approx-no-cov} into an estimate, we plug in the empirical variances \(\widehat{\operatorname{Var}}(N_i)\) and \(\widehat{\operatorname{Var}}(D_j)\), that is, the sample variances of the sequences \(\{N_i\}_{i=1}^{S_2}\) and \(\{D_j\}_{j=1}^{S_1}\) with divisors \(S_2 - 1\) and \(S_1 - 1\), and the effective sample size \(S_{1,\mathrm{eff}}\) estimated from the sequence \(\{D_j\}\). We denote the resulting plug-in estimate by \(\widehat{\operatorname{Var}}(\widehat{Z})\). The MCSE is then:
\begin{equation*}
\operatorname{MCSE}\left(\widehat{Z}\right) = \sqrt{ \widehat{\operatorname{Var}}\left( \widehat{Z}\right)}.
\end{equation*}
Evaluated at the fixed point of the iterated bridge function, the relative variance in Equation \eqref{eq:var-approx-no-cov} recovers the relative mean-squared error approximation of \citet{fruhwirth-schnatter_estimating_2004}, which builds on the asymptotic variance of \citet{meng_simulating_1996}. Our formulation differs in two respects. First, the autocorrelation correction uses the multi-chain effective sample size of \citet{Vehtari+etal:2021:Rhat} instead of a single-sequence autoregressive spectral estimate. Second, the variance is propagated to the log scale via the exact log-normal identity below, instead of the first-order equivalence $E(\log \widehat{Z} - \log Z)^2 \approx \operatorname{RE}^2$ of \citet[eq.~14]{fruhwirth-schnatter_estimating_2004}. The two agree when $\operatorname{RE}^2$ is small, but differ in the saturation regime of Section \ref{sec:saturation-cap}.

To compute the MCSE of the logarithm of the marginal likelihood estimate, \(\log \widehat{Z}\), we use the property of the log-normal distribution. The theoretical variance is given by:
\begin{equation*}
\operatorname{Var}\left( \log \widehat{Z} \right) = \log\left( 1 + \frac{\operatorname{Var}\left( \widehat{Z} \right)}{Z^2} \right).
\end{equation*}
By substituting the point estimate \(\widehat{Z}\) for the true normalizing constant \(Z\), and using our variance approximation from Equation \eqref{eq:var-approx-no-cov}, we obtain the estimated variance:
\begin{equation*}
\widehat{\operatorname{Var}}\left( \log \widehat{Z} \right) = \log\left( 1 + \frac{\widehat{\operatorname{Var}}\left( \widehat{Z} \right)}{\widehat{Z}^2} \right).
\end{equation*}
Then
\begin{equation*}
    \operatorname{MCSE}\left(\log \widehat{Z}\right) =
    \sqrt{ \widehat{\operatorname{Var}}\left( \log \widehat{Z}\right)}.
\end{equation*}
The MCSE quantifies the uncertainty in the estimated (log) marginal likelihood due to the finite number of draws $S_1$ and $S_2$, conditional on the proposal and the final bridge function when the iteration ends. However, the standard MCSE is only well-calibrated when the central limit theorem holds. In Section \ref{sec:saturation-cap}, we establish the empirical saturation cap of this estimator to diagnose pre-asymptotic failures. 

\section{Saturation Cap and Pre-asymptotic Behavior}
\label{sec:saturation-cap}

The derivations in Section \ref{sec:mcse} rely on the assumption that the estimator operates in an asymptotic regime, where $S_1$ and $S_2$ are sufficiently large for the central limit theorem (CLT) to hold. In the pre-asymptotic regime, often caused by high dimensionality or severe tail mismatch between the posterior and the proposal, a small number of draws can dominate the bridge weights. In this state the approximations fail, and the MCSE underestimates the true variability. Both the central limit theorem and the second-order treatment of the shared bridge function in Section \ref{sec:mcse} rely on the asymptotic regime.

The magnitude of the MCSE estimate serves directly as an indicator of this pre-asymptotic failure. To formalize this, we decompose the relative variance of the estimator additively as
\begin{equation}
\label{eq:variance-decomposition}
\frac{\widehat{\operatorname{Var}}(\widehat{Z})}{\widehat{Z}^2} \approx T_N + T_D,
\end{equation}
where the proposal and MCMC terms are given by
\[
T_N \coloneqq \frac{\widehat{\operatorname{Var}}(N_i)}{S_2 \bar{N}^2}
\quad \text{and} \quad
T_D \coloneqq \frac{\widehat{\operatorname{Var}}(D_j)}{S_{1,\mathrm{eff}} \bar{D}^2}.
\]
Consider a sample of $S$ draws of a non-negative random variable $X$. Its empirical relative variance $\operatorname{Var}(X)/\bar{X}^2$ is bounded by $S - 1$ under the $1/S$ variance convention, and by $S$ under the unbiased $1/(S-1)$ convention used for the plug-in estimates in Section \ref{sec:mcse}. Either inequality saturates when a single observation carries all the empirical weight, and the difference is immaterial at the sample sizes considered here. Treating the ESS-corrected denominator as a set of $S_{1,\mathrm{eff}}$ effectively i.i.d.\ draws, these structural bounds apply to the per-mean forms, yielding $T_N \lesssim 1$ and $T_D \lesssim 1$. 

Consequently, the total relative variance encounters an empirical ceiling:
\[
\frac{\widehat{\operatorname{Var}}(\widehat{Z})}{\widehat{Z}^2} \lesssim 2.
\]
Applying this to the log-normal variance identity gives an upper bound on the estimated variance of the log marginal likelihood:
\[
\widehat{\operatorname{Var}}\left( \log \widehat{Z} \right) = \log\left( 1 + \frac{\widehat{\operatorname{Var}}(\widehat{Z})}{\widehat{Z}^2} \right) \lesssim \log(1 + 2) = \log 3.
\]
Therefore, the standard error is capped at:
\begin{equation}
\label{eq:mcse-cap}
\operatorname{MCSE}\left(\log \widehat{Z}\right) \lesssim \sqrt{\log 3} \approx 1.05.
\end{equation}
The cap is reached in stages. If a single draw carries nearly all the empirical weight in one of the two sums, that term alone saturates ($T_N \approx 1$ or $T_D \approx 1$), giving $\operatorname{MCSE}(\log \widehat{Z}) \approx \sqrt{\log 2} \approx 0.83$. The full cap requires both terms to saturate at once. When the empirical MCSE approaches these values, it ceases to be a calibrated estimate of uncertainty: the reported value signals saturation, not precision, and the true error can be much larger. For the `State-votes hier-GP' posterior in Section \ref{sec:results}, for example, the cross-repetition standard deviation is $\approx 4.8$ while the MCSE reads $\approx 1.0$. Calibration degrades well before the cap is reached, and based on the extensive experimental results in Section \ref{sec:results} we recommend the practical trust threshold $\operatorname{MCSE}\left(\log \widehat{Z}\right) < 0.3$.

The failure mode formalized above is a few draws dominating the summand, so that CLT-based error estimates become uncalibrated. The Pareto-$\hat{k}$ diagnostic \citep{vehtari_pareto_2024} targets the same failure mode, fitting a generalized Pareto distribution (GPD) to the largest importance ratios and flagging $\hat{k} > 0.7$. We evaluated Pareto-$\hat{k}$ on the bridge summands $N_i$ and $D_j$ for all posteriors and repetitions in Section \ref{sec:results}. The GPD did not describe the summand tails well: $\hat{k}$ estimates were high and sensitive to the number of tail draws used in the fit. Empirically the 0.7 threshold based on the convergence rate of Pareto tailed summands is not a good diagnostic threshol for the bridge summands, leaving no justified cutoff. An extended log-normal tail model \citep{Beirlant+Bladt:2025} matched the summand tails better, but diagnostic based on that did not improve on the MCSE threshold. The bridge sampling MCSE is driven primarily by tail \emph{scale}. Across our corpora the strongest single predictor of the per-repetition MCSE was the log-normal tail-scale parameter, not either shape diagnostic. Since the MCSE is both the quantity of interest and directly calibrated against ground truth (Section \ref{sec:results}), we use it with the 0.3 threshold as the single decision rule.

\section{Hybrid Score-Matching for Proposal Distributions}
\label{sec:hybrid-score-matching}

The standard multivariate normal proposal for bridge sampling, $g(\vectheta) = \mathcal{N}(\hat{\boldsymbol{\mu}}, \hat{\boldsymbol{\Sigma}}_n)$, relies heavily on the sample covariance $\hat{\boldsymbol{\Sigma}}_n$. This formulation is highly sensitive to operator-norm errors. For i.i.d.\ sub-Gaussian samples, the effective-rank bound \citep{koltchinskii_inequalities_2014} scales as:
\begin{equation*}
\|\hat{\boldsymbol{\Sigma}}_n - \boldsymbol{\Sigma}\|_{\mathrm{op}} \;\lesssim\; \|\boldsymbol{\Sigma}\|_{\mathrm{op}} \sqrt{\frac{\mathbf{r}(\boldsymbol{\Sigma})}{S_{\mathrm{fit}}}} \;+\; \|\boldsymbol{\Sigma}\|_{\mathrm{op}} \frac{\mathbf{r}(\boldsymbol{\Sigma})}{S_{\mathrm{fit}}},
\end{equation*}
where $\mathbf{r}(\boldsymbol{\Sigma}) \coloneqq \operatorname{tr}(\boldsymbol{\Sigma})/\|\boldsymbol{\Sigma}\|_{\mathrm{op}} \le p$ is the effective rank, and $S_{\mathrm{fit}}$ is the number of posterior draws used to fit the proposal. We use this bound heuristically for autocorrelated MCMC draws, replacing $S_{\mathrm{fit}}$ with its effective sample size. Heavy-tailed targets such as the horseshoe posteriors in Section \ref{sec:results} fall outside the theorem's hypotheses, so the bound serves as motivation for the scaling rather than a guarantee. As the ratio $\mathbf{r}(\boldsymbol{\Sigma})/S_{\mathrm{fit}}$ grows, $\hat{\boldsymbol{\Sigma}}_n$ becomes increasingly misaligned with the true posterior covariance $\boldsymbol{\Sigma}$. Consequently, the tail of $\widetilde{\pi}/g$ becomes heavy, and in our experiments the bridge sampling estimator becomes unreliable well before $\mathbf{r}(\boldsymbol{\Sigma})$ approaches $S_{\mathrm{fit}}$ (Section \ref{sec:results}). Warping does not resolve this regime: moment-matching warp transformations standardize the draws using the same estimate $\hat{\boldsymbol{\Sigma}}_n$ and therefore inherit its spectral error. Consistent with this, in our experiments Warp-III bridge sampling gave only modest improvements where the sample proposal was already reliable, but degraded or failed exactly where $\mathbf{r}(\boldsymbol{\Sigma})/S_{\mathrm{fit}}$ is largest (Section \ref{sec:results}).

To regularize the proposal covariance in this moderate-effective-rank regime, we incorporate a second source of information: the local geometry of the posterior captured by its gradients. Let 
\[
s_i \coloneqq \nabla\log\widetilde{\pi}(\vectheta_i)
\]
denote the scores evaluated at the MCMC draws. The centered empirical second moment of the scores,
\begin{equation*}
\hat{\mathbf{I}} \;=\; \frac{1}{S_{\mathrm{fit}}}\sum_{i=1}^{S_{\mathrm{fit}}} (s_i - \bar{s})(s_i - \bar{s})^\top,
\end{equation*}
is a consistent estimator of the Fisher information matrix $I(\pi) = \mathbb{E}_\pi[s s^\top] = -\mathbb{E}_\pi[\nabla^2\log\pi]$. For a perfectly Gaussian target, $I(\pi) = \boldsymbol{\Sigma}^{-1}$, meaning $\hat{\boldsymbol{\Sigma}}_{\mathrm{score}} \coloneqq \hat{\mathbf{I}}^{-1}$ provides a consistent estimator of the covariance. For non-Gaussian posteriors, $\hat{\boldsymbol{\Sigma}}_{\mathrm{score}}$ targets $I(\pi)^{-1}$, the inverse of the average negative log-posterior curvature. These two targets are not the same. The Cram\'{e}r--Rao inequality applied to the location family generated by $\pi$ gives $\boldsymbol{\Sigma} \succeq I(\pi)^{-1}$, with equality if and only if $\pi$ is Gaussian. For non-Gaussian targets $\hat{\boldsymbol{\Sigma}}_{\mathrm{score}}$ is therefore systematically narrower than $\boldsymbol{\Sigma}$ in the positive-semidefinite order. On its own it would yield a too-narrow proposal, and hence a heavy tail in the importance ratio $\widetilde{\pi}/g$. We use it only as a low-variance regularizing component, not as a replacement for $\hat{\boldsymbol{\Sigma}}_n$.

We combine the sample covariance $\hat{\boldsymbol{\Sigma}}_n$ and the score-based covariance $\hat{\boldsymbol{\Sigma}}_{\mathrm{score}}$ using the affine-invariant Riemannian metric (AIRM) midpoint on the positive-definite manifold:
\begin{equation*}
\hat{\boldsymbol{\Sigma}}_{\mathrm{hybrid}}
\;=\; \hat{\boldsymbol{\Sigma}}_n \,\#\, \hat{\boldsymbol{\Sigma}}_{\mathrm{score}}
\;=\; \hat{\boldsymbol{\Sigma}}_n^{1/2}
\left(\hat{\boldsymbol{\Sigma}}_n^{-1/2}\,\hat{\boldsymbol{\Sigma}}_{\mathrm{score}}\,
\hat{\boldsymbol{\Sigma}}_n^{-1/2}\right)^{1/2}
\hat{\boldsymbol{\Sigma}}_n^{1/2}.
\end{equation*}
\citet[Theorem 2.3]{seyboldt_preconditioning_2026} show that this midpoint of the draw covariance and the inverse score covariance minimizes the empirical Fisher divergence between the target posterior $\pi$ and the Gaussian proposal. The result holds for the finite-sample objects themselves and does not assume posterior Gaussianity. The narrowing of $\hat{\boldsymbol{\Sigma}}_{\mathrm{score}}$ is thus balanced against the spectral noise of $\hat{\boldsymbol{\Sigma}}_n$ by an explicit optimality criterion rather than by a tuning parameter. The narrowing is also only partly inherited: when $\hat{\boldsymbol{\Sigma}}_{\mathrm{score}} \preceq \hat{\boldsymbol{\Sigma}}_n$, the midpoint satisfies $\hat{\boldsymbol{\Sigma}}_{\mathrm{score}} \preceq \hat{\boldsymbol{\Sigma}}_{\mathrm{hybrid}} \preceq \hat{\boldsymbol{\Sigma}}_n$. This optimality concerns gradient matching, and was derived for HMC mass-matrix adaptation rather than for proposal construction. Section \ref{sec:results} shows empirically that it also greatly improves the proposal distribution for bridge sampling. 

In the hybrid bridge sampling workflow, $\hat{\boldsymbol{\Sigma}}_{\mathrm{hybrid}}$ directly replaces $\hat{\boldsymbol{\Sigma}}_n$ in the proposal density $g(\vectheta) = \mathcal{N}(\hat{\boldsymbol{\mu}}, \hat{\boldsymbol{\Sigma}}_{\mathrm{hybrid}})$, while the proposal mean remains the sample mean $\hat{\boldsymbol{\mu}} = \bar{\vectheta}$. The fundamental bridge identity and the iterative optimal-bridge updates remain unchanged.

The scores are computed on the same fitting subset used for $\hat{\boldsymbol{\mu}}$ and $\hat{\boldsymbol{\Sigma}}_n$, in the unconstrained parameter space.
The extra computational cost is one gradient evaluation per fitting draw and an $O(p^3)$ eigendecomposition for the AIRM matrix square roots, both small relative to the cost of generating the MCMC draws for $p \lesssim 10^3$ and $S_{\mathrm{fit}} \lesssim 10^4$. The AIRM midpoint requires both inputs to be strictly positive definite. All experiments in Section \ref{sec:results} have $p < S_{\mathrm{fit}}$, so $\hat{\boldsymbol{\Sigma}}_n$ and $\hat{\mathbf{I}}$ are full-rank. As a defensive measure against numerical round-off, we project both $\hat{\boldsymbol{\Sigma}}_n$ and $\hat{\boldsymbol{\Sigma}}_{\mathrm{score}}$ to the nearest positive-definite matrix by clipping small negative eigenvalues before combining. We make no claim about the rank-deficient regime $S_{\mathrm{fit}} \le p$. When the hybrid proposal is nevertheless badly mismatched, the diagnostics indicate the failure. Either the bridge iteration produces non-finite ratios and aborts, or the MCSE approaches the saturation cap of Section \ref{sec:saturation-cap}. The `State-votes hier-GP' posterior in Section \ref{sec:results} is an example of this.

\section{Experimental Results}
\label{sec:results}

This section evaluates the empirical calibration of the bridge sampling MCSE and assesses the estimator's precision across increasingly challenging geometries. 
Analytical ground truth is unavailable for the complex posteriors in \texttt{posteriordb}, so we establish a computational reference (``gold standard'') for each posterior. We run independent MCMC and bridge sampling repetitions with a high number of posterior draws, and average their log marginal likelihood estimates on the log scale. We use 200 repetitions for the horseshoe experiments and 100 for \texttt{posteriordb}. This average is the proxy for the true log marginal likelihood against which the MCSE estimates are validated.
Throughout, a total posterior sample of size $S$ is divided by a 50/50 split into a fitting subset of size $S/2$, used to fit the proposal $g(\vectheta)$, and an evaluation subset of size $S_1 = S/2$; the number of proposal draws is set equal, $S_2 = S_1$ \citep{overstall_forster_2010}.
In both experiments each repetition is evaluated at two total sample sizes $S \in \{4000, 16000\}$ using nested draws: the $S = 4000$ evaluation uses a subset of the same posterior draws as the $S = 16000$ evaluation. The nesting is intentional (a common-random-numbers design): it removes independent MCMC noise from within-repetition comparisons across $S$, so differences between the two sample sizes reflect the effect of the sample size rather than chain-to-chain variation. All reported aggregates (RMSE, MCSE, coverage) are computed per posterior for a single combination of $S$ and proposal, where the repetitions remain mutually independent.
We evaluate both the standard sample-covariance multivariate normal proposal (`sample') and the regularized AIRM-midpoint preconditioner (`hybrid'). We also evaluated Warp-III bridge sampling \citep{meng_schilling_2002, gronau_computing_2020} on all horseshoe and \texttt{posteriordb} posteriors with the same repetitions. Warp-III reduced the cross-repetition standard deviation by a modest 15--20\% for the posteriors and sample sizes where the `sample' proposal was already reliable, but at $S = 4000$ it became worse than the `sample' proposal for $k \in \{80, 90, 100\}$ and its proposal collapsed on the largest \texttt{posteriordb} models; the `hybrid' proposal's improvements are several times larger throughout. We therefore do not report the warp results separately. Our analysis focuses primarily on the precision of the estimates, as we utilize the standard bridge sampling estimator without bias-inducing modifications, meaning the estimator remains consistent and asymptotically unbiased.

\subsection{Horseshoe Regression with Increasing Dimension}

We begin by exploring the performance of bridge sampling within the context of linear regression with a highly non-normal posterior, particularly focusing on how the method behaves as the number of covariates increases and the posterior gets higher dimensional.

The simulated data for this analysis were generated as follows. For $n = 100$ observations, the design matrix $\mathbf{X}$ was created with an increasing number of $k\in\{10,20,\dots,100\}$ covariates as follows:
\[
\mathbf{X} = \frac{1}{\sqrt{n \times k}} \boldsymbol{\xi},
\]
where $\boldsymbol{\xi} \in \mathbb{R}^{n \times k}$, with $\boldsymbol{\xi}_{ij} \overset{\text{i.i.d.}}{\sim} \mathcal{N}(0, 1)$.

The regression coefficients $\boldsymbol{\beta}$ were sampled from a standard normal distribution, with the last coefficient set to zero:

\[
\boldsymbol{\beta} = (\beta_1, \beta_2, \dots, \beta_{k-1}, 0)^\top, \quad \text{where } \beta_i \sim \mathcal{N}(0, 1).
\]
The response variable $\mathbf{y}$ was generated as a linear combination of the covariates with added normal noise:

\[
\mathbf{y} = \mathbf{X} \boldsymbol{\beta} + \frac{1}{\sqrt{n}} \boldsymbol{\varepsilon},
\]
where $\boldsymbol{\varepsilon}=\{\varepsilon_1,\dots,\varepsilon_n\}$, with $\varepsilon_i\sim \mathcal{N}(0, 2)$. 

To induce a non-normal posterior, a regularized horseshoe prior $\text{RHS}(\tau, \lambda_i)$ \citep{Piironen+Vehtari:2017:rhs} was used for the regression coefficients $\boldsymbol{\beta}$. The RHS prior was implemented as a scale mixture of normals with slab-regularized local scales:
\[
\beta_i \sim \mathcal{N}(0, \tau^2 \tilde{\lambda}_i^2), \quad
\tilde{\lambda}_i^2 = \frac{c^2 \lambda_i^2}{c^2 + \tau^2 \lambda_i^2}, \quad
\lambda_i \sim \text{Cauchy}^+(0, 1), \quad \tau \sim \text{Cauchy}^+(0, \tau_0 \sigma),
\]
where the slab scale is $c^2 = s_{\mathrm{slab}}^2\, c_{\mathrm{aux}}$ with $c_{\mathrm{aux}} \sim \text{Inv-Gamma}(\nu_{\mathrm{slab}}/2,\, \nu_{\mathrm{slab}}/2)$. We use the \texttt{rstanarm} \texttt{hs()} defaults $\tau_0 = 0.01$, $s_{\mathrm{slab}} = 2.5$, $\nu_{\mathrm{slab}} = 4$, and the \texttt{rstanarm} default priors for the intercept $\beta_0$ and the residual scale $\sigma$.
This mismatch between the data generation (Normal) and the estimation model (RHS) is intentional. It creates a challenging, target distribution that becomes strongly non-Gaussian in the scale-mixture hyperparameters. \texttt{rstanarm} samples this model in a non-centered parameterization. Each covariate contributes three unconstrained parameters: a raw coefficient $z_i$, from which $\beta_i = z_i\, \tau \tilde{\lambda}_i$ is a deterministic transform, and two parameters decomposing the half-Cauchy $\lambda_i$ into a normal and an inverse-gamma component. Five parameters are global: two decomposing $\tau$ in the same way, the slab parameter $c_{\mathrm{aux}}$, the intercept $\beta_0$, and the residual scale $\sigma$. The unconstrained posterior dimension is therefore $p = 3k + 5$.

The posterior sampling was conducted using the \texttt{rstanarm} function \texttt{stan\_glm} \citep{rstanarm} with bridge sampling via the \texttt{bridgesampling} package \citep{gronau_bridgesampling_2020}.
For each $k$ a single dataset is generated, and 1000 independent evaluation repetitions vary only the MCMC and Pathfinder seeds, so the reported variability is that of the estimator for a fixed dataset. Each repetition obtains $16000$ post-warmup draws (4 chains $\times$ 4000), and bridge sampling is evaluated at total posterior sample sizes $S = 16000$ (the full pool) and $S = 4000$ (subsampled via \texttt{posterior::subset\_draws()}), under both proposals.
The reference always averages 200 independent repetitions with the hybrid proposal. For $k \in \{10, 20, \dots, 70\}$ these use $S = 32000$ (4 chains $\times$ 8000 post-warmup). For $k \in \{80, 90, 100\}$ the increasingly non-Gaussian posterior makes the bridge sampling variance much larger, and the repetitions use $S = 128000$ (32 chains $\times$ 4000 post-warmup). The reference posterior
sample sizes were chosen so that based on increasing $S$ the averaged estimate and standard deviation behaved eventually according to central limit theorem and the estimates were clearly stable and with small 
variation.

\begin{figure}[tp]
  \centering
  \includegraphics[width=\textwidth]{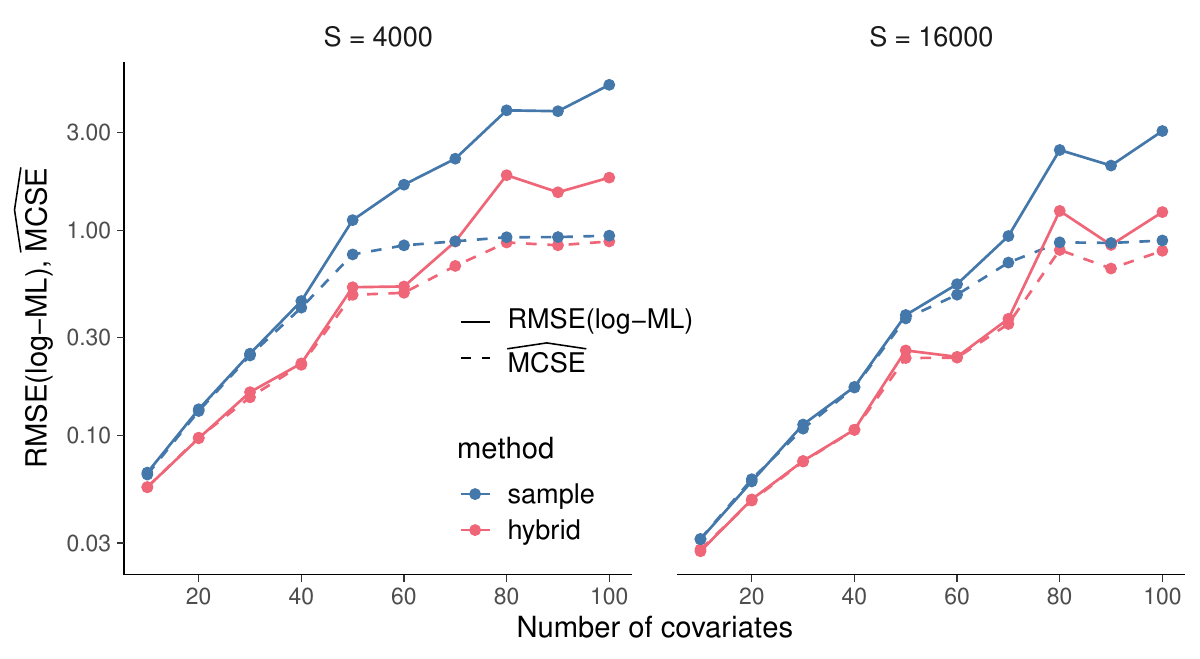}
  \caption{Root-mean-square error $\mathrm{RMSE}(\log\text{-ML})$ with respect to the independent high-$S$ reference and the empirical $\widehat{\mathrm{MCSE}}$, comparing the `sample' and `hybrid' proposals across $k \in \{10, \dots, 100\}$ covariates at $S \in \{4000, 16000\}$, with 1000 independent evaluation repetitions for each $k$. Lines where RMSE coincides with MCSE indicate the bridge sampler is well calibrated.}
  \label{fig:hs_hybrid}
\end{figure}

Figure \ref{fig:hs_hybrid} shows the estimated MCSE and the root mean
square error (RMSE) against the reference values using 1000
independent repetitions. The standard `sample' proposal degrades as
$k$ increases: at $S = 4000$ its RMSE exceeds the estimated MCSE by a
factor of 2.0 at $k = 60$, growing to 5.4 at $k = 100$, indicating
uncaptured pre-asymptotic variability and residual bias. The `hybrid'
score-matching proposal reduces the RMSE by a factor of 1.5--3 across
$k \ge 40$, with the largest reduction of $3.1\times$ at $k = 60$,
$S = 4000$. It also keeps the empirical RMSE within 10\% of the
analytical $\widehat{\mathrm{MCSE}}$ up to $k = 60$ at $S = 4000$, and
up to $k = 70$ at $S = 16000$. The hybrid covariance thus rescues the
estimator's calibration in these moderately high dimensions
($p = 3k + 5$ up to $\approx 220$) without requiring an increase in
MCMC draws. From $k = 80$ ($p = 245$) onward the hybrid estimates
degrade as well, with RMSE/MCSE reaching 1.3--2.1, and a higher
posterior sample size would be needed.

\subsection{\texorpdfstring{\texttt{posteriordb}}{posteriordb} Posteriors}

To validate these findings on real-world geometries and gain further
insights into the performance of bridge sampling under different
conditions, we evaluate 41 models and posteriors from the
\texttt{posteriordb} database \citep{Magnusson-etal:2025:posteriordb}
plus a multi-component Gaussian process model of birthday data
\citep{zhang_pathfinder_2022}. These span $p = 1$ to $p = 1006$
parameters, including a representative set of posteriors with varying
dimensionality, non-normality, funnelness, and multi-modality.

Each posterior was sampled 100 times using \texttt{CmdStanR} with the
default options and the Pathfinder algorithm
\citep{zhang_pathfinder_2022} to initialize MCMC for improved
convergence speed and reliability. Each repetition yielded 4000 and
16000 post-warmup draws (4 chains $\times$ 1000 and 4 chains $\times$ 4000,
respectively), and bridge sampling was evaluated at both
$S \in \{4000, 16000\}$ under both proposals, giving four bridge
estimates per repetition.
For 39 of the 42 posteriors we used 100 independent repetitions at
$S = 32000$ (4 chains $\times$ 8000 post-warmup). For `SAT Rasch' and
`LSAT Rasch' we used $S = 64000$ (32 chains $\times$ 2000 post-warmup). The reference posterior
sample sizes were chosen so that based on increasing $S$ the averaged
estimate and standard deviation behaved eventually according to
central limit theorem and the estimates were clearly stable and with
small variation. For the remaining model,
`State-votes hier-GP' ($p = 933$), we were not able to compute a trustworthy
reference, and it is therefore excluded from the calibration analysis. The
hybrid proposal produced too large log importance ratios, leading to
underflow and bridge iteration failure. The sample proposal produced
finite estimates at every tested $16000 \leq S \leq 128000$, but with
mean drifting up (from 2443 to 2450), standard deviation over repetitions shrinking slowly
(from 4.8 to 3.3), and $\widehat{\mathrm{MCSE}}$ near the cap (from 1.0
to 0.9), the estimates were not reliable. These numbers also
illustrate the saturation behavior of Section \ref{sec:saturation-cap}
directly: the MCSE, pinned near its cap, understates the observed
cross-repetition standard deviation four- to five-fold.

\begin{figure}[tp]
  \centering
  \includegraphics[width=0.95\textwidth]{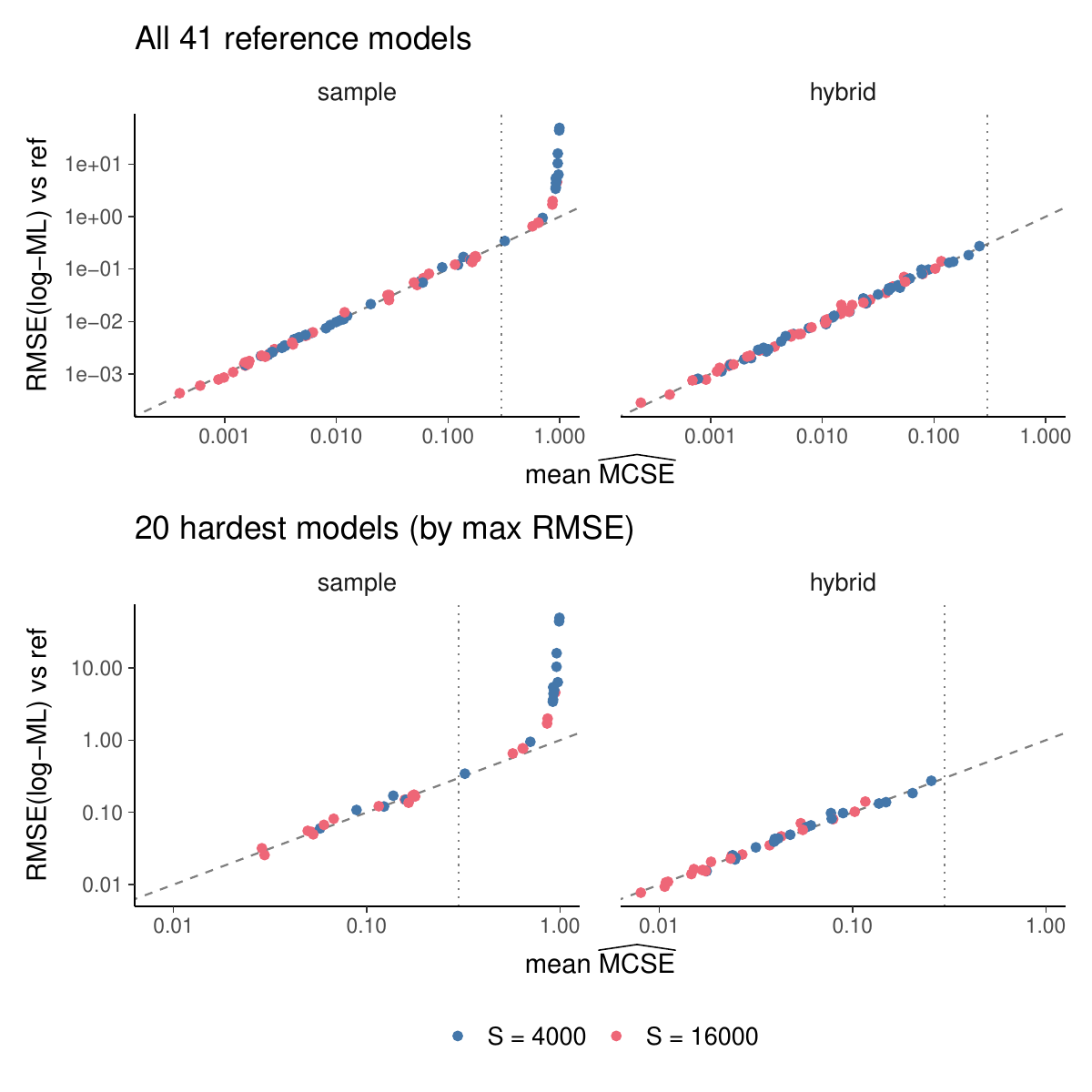}
  \caption{Estimated $\widehat{\mathrm{MCSE}}$ vs. $\mathrm{RMSE}(\log\text{-ML})$, computed from 100 independent evaluation repetitions against the independent high-$S$ reference values, for the 41 analyzed posteriors (40 \texttt{posteriordb} models plus the Birthdays GP). The top row displays all models, while the bottom row isolates the 20 hardest models by max RMSE. Points on the $y = x$ diagonal indicate that the MCSE accounts for the total error, which requires both negligible bias in $\log \widehat{Z}$ and an accurate MCSE estimate of the sampling standard deviation. Vertical dotted lines mark $\widehat{\mathrm{MCSE}} = 0.3$. The MCSE saturation cap is visible in the left-side subplots.}
  \label{fig:pdb_scatter}
\end{figure}

\begin{figure}[tp]
  \centering
  \includegraphics[width=\textwidth]{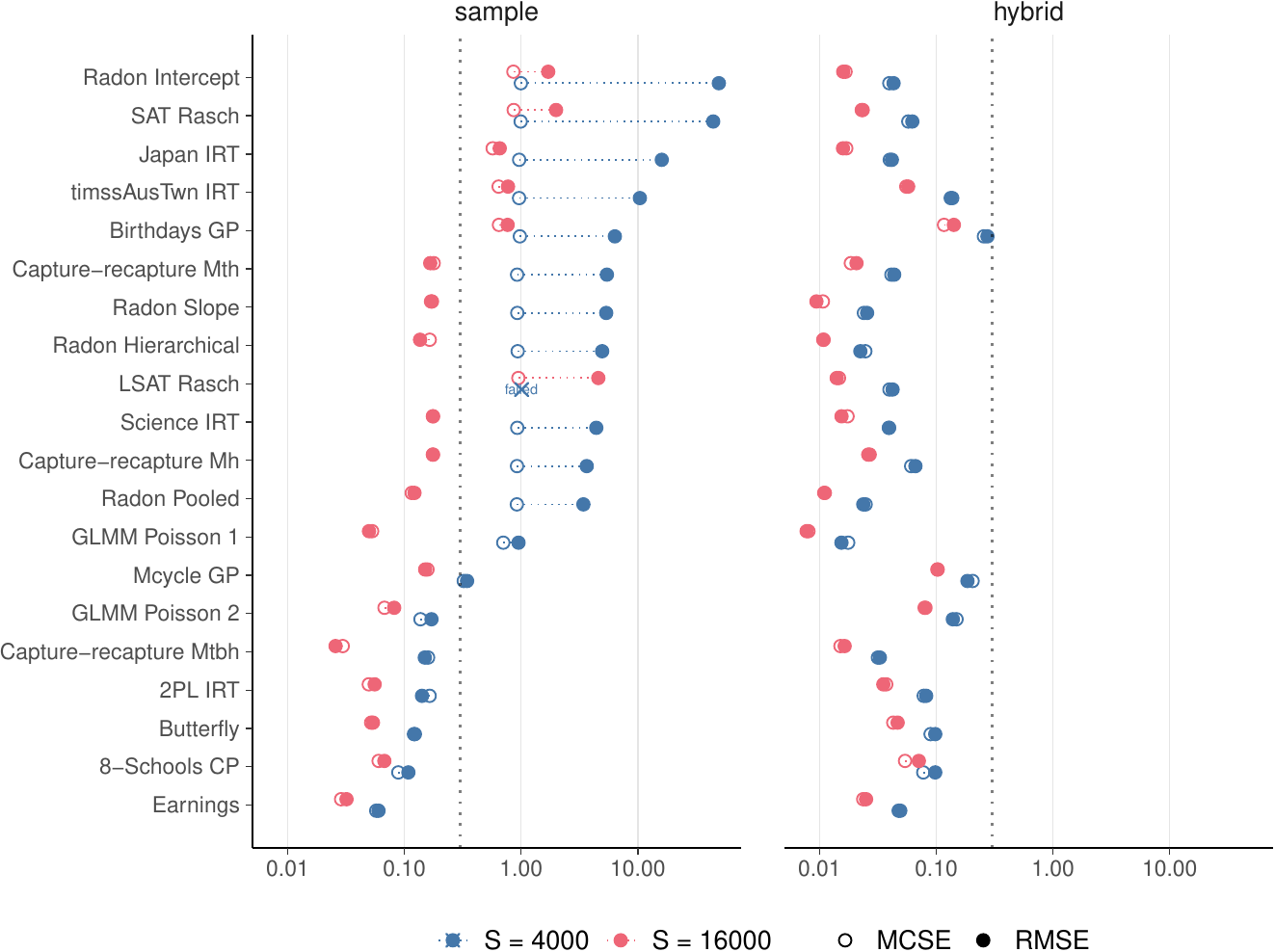}
  \caption{Dumbbell plot for the 20 hardest \texttt{posteriordb} posteriors. Filled circles denote $\mathrm{RMSE}(\log\text{-ML})$ against the reference, and open circles denote $\widehat{\mathrm{MCSE}}$, connected by a dotted segment. Short segments indicate the MCSE faithfully summarizes the total error. Vertical dotted lines mark $\widehat{\mathrm{MCSE}} = 0.3$. The MCSE saturation cap at $\approx 1.05$ is visible on the left side subplot.}
  \label{fig:pdb_dumbbell}
\end{figure}

Failed repetitions are excluded from all RMSE and coverage aggregates,
with the following accounting. A repetition is successful if the
bridge sampler returned a finite log-ML without aborting. Of the
$16{,}760$ candidate evaluation repetitions (42 models $\times$ 100
repetitions $\times$ 2 proposals $\times$ 2 values of $S$), 402
($2.4\,\%$) fail this criterion. The failures concentrate in two
posteriors. The first is `LSAT Rasch' under the `sample' proposal at
$S = 4000$, where all 100 repetitions fail; no RMSE is reported for it
and the combination is annotated ``failed'' in Figures
\ref{fig:pdb_scatter} and \ref{fig:pdb_dumbbell}. The second is
`State-votes hier-GP', which is excluded entirely as described above. An
MCMC-convergence filter (more than $10\,\%$ post-warmup divergences or
lp\_\_ ESS below 20), applied symmetrically to reference and
evaluation repetitions, drops a further 46 of the remaining $16{,}358$
($0.28\,\%$). In the horseshoe experiment no repetitions were dropped
by either criterion. For posteriors where only some repetitions fail, the RMSE is therefore
conditional on bridge-sampling success. Outside the two posteriors named
above, the proportion of failures is negligible. We apply the same
filters to the reference and the evaluation repetitions, so that neither
side of the comparison is selectively cleaned.

Figure \ref{fig:pdb_scatter} shows RMSE and estimated MCSE based on
100 independent repetitions for each posterior. At $S = 16000$ the
`hybrid' proposal lies on the parity diagonal for all
41 models with a reference: $\mathrm{RMSE}/\widehat{\mathrm{MCSE}} \le
1.4$ for every model, and $\le 1.2$ for 38 of them. The `sample'
proposal is comparably calibrated for 38 of the 41 models, with
`LSAT Rasch' the extreme exception at
$\mathrm{RMSE}/\widehat{\mathrm{MCSE}} = 4.8$. The analytic MCSE is thus
a faithful summary of the total error.
Since $\mathrm{RMSE}^2 = \mathrm{bias}^2 + \mathrm{SD}^2$, lying on the
diagonal requires two things: negligible estimator bias, and an accurate
MCSE estimate of the sampling standard deviation. Both hold at
$S = 16000$. Across models and proposals, the median
$|\mathrm{bias}|/\mathrm{RMSE}$ is 0.08, and the median ratio of
cross-repetition SD to mean MCSE is 1.0. At $S = 4000$ the gap widens: under the `sample' proposal,
11 of 40 models drift above the diagonal with
$\mathrm{RMSE}/\widehat{\mathrm{MCSE}} > 2$ (up to 50 for `Radon
Intercept' and 45 for `SAT Rasch'), and `LSAT Rasch' fails
outright. This drift is bias-dominated: bias accounts for 61--98\% of
the RMSE in 10 of the 11 models, while the cross-repetition SD (2--10)
also far exceeds the near-cap MCSE ($\approx 0.9$--$1.0$). Under the
`hybrid' proposal every model remains within
$\mathrm{RMSE}/\widehat{\mathrm{MCSE}} \le 1.3$, and on the 11 models
miscalibrated under the `sample' proposal the RMSE is reduced by
factors of 23 to over 1000.

Figure \ref{fig:pdb_dumbbell} zooms in on the 20 hardest posteriors. The most extreme failure occurs in the \texttt{lsat} model ($p = 1006$). At a 50/50 split of $S=4000$, the fit subset contains $S_{\mathrm{fit}} = 2000$ draws. The ratio $p/S_{\mathrm{fit}} \approx 0.50$ triggers severe spectral error in the sample covariance, causing the bridge iteration to abort immediately (marked as ``failed'' for the `sample' method). The `hybrid' combiner successfully stabilizes the covariance matrix, yielding useful estimates. 

\subsection{MCSE Calibration and Coverage}

Finally, we assess the empirical calibration of the $\widehat{\mathrm{MCSE}}$. The per-repetition standardized error is defined as $z = (\log\hat Z - \log\hat Z_{\text{ref}})/\widehat{\mathrm{MCSE}}$, with a miscoverage event of the 95\% CI occurring when $|z| > 1.96$. By pooling the horseshoe and \texttt{posteriordb} replicates that pass the success and MCMC-convergence filters of Section \ref{sec:results} and partitioning them into 30 equal-count bins by $\widehat{\mathrm{MCSE}}$ magnitude, we track the empirical coverage of the 95\% confidence interval against the high-$S$ reference values. 

\begin{figure}[tp]
  \centering
  \includegraphics[width=0.85\textwidth]{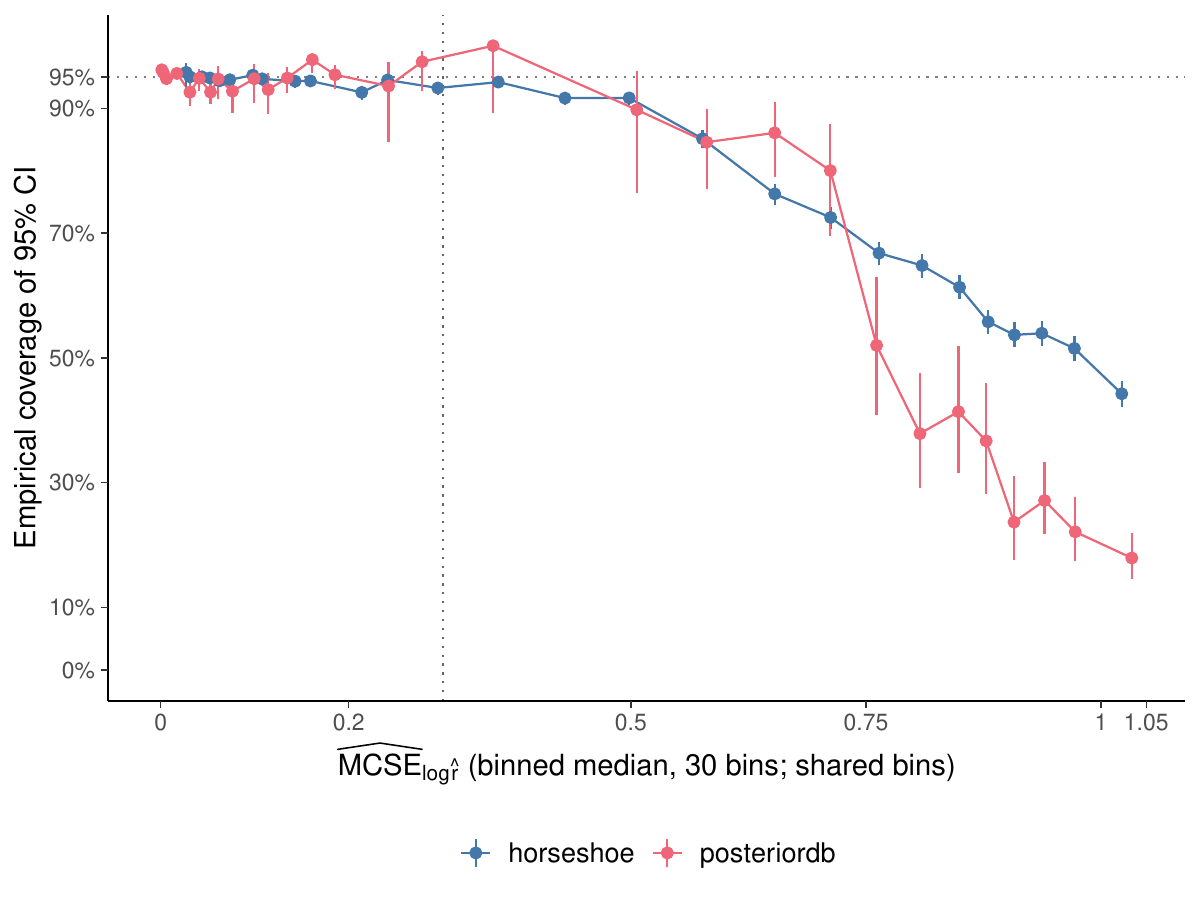}
  \caption{Empirical coverage of the 95\% confidence intervals across 30 equal-count bins of $\widehat{\mathrm{MCSE}}$, pooling the horseshoe and \texttt{posteriordb} corpora. Coverage is well calibrated below the 0.3 threshold (vertical dotted line) and smoothly degrades as it approaches the theoretical $\sqrt{\log 3} \approx 1.05$ saturation cap.}
  \label{fig:mcse_calib}
\end{figure}

As shown in Figure \ref{fig:mcse_calib}, the empirical coverage is well calibrated at $\widehat{\mathrm{MCSE}} < 0.3$: across the 29 bins below the threshold, coverage ranges from 0.93 to 0.98 with median 0.95. Above the threshold the coverage degrades steadily as the estimate approaches the $\approx 1.05$ saturation cap, falling to 0.18--0.44 in the bins with $\widehat{\mathrm{MCSE}} \approx 1$. Note that the MCSE conditions on the fitted proposal (Section \ref{sec:mcse}), whereas the cross-repetition coverage and RMSE include the variability of refitting the proposal on fresh draws; the observed calibration therefore implies that proposal-refit variability is negligible below the 0.3 threshold. This justifies the diagnostic rule. Practitioners can trust $\widehat{\mathrm{MCSE}}$ when it falls below 0.3. Above the threshold, either a larger $S$ or a better proposal, such as the hybrid AIRM midpoint, is needed.

\section{Discussion and Conclusion}
\label{sec:discussion}
This paper introduces a combined methodological and diagnostic framework to improve and quantify the reliability of marginal likelihood estimation via bridge sampling. 

First, we presented and validated a closed-form Monte Carlo standard error (MCSE) for the estimator, building on the variance approximations of \citet{meng_simulating_1996} and \citet{fruhwirth-schnatter_estimating_2004}. Our estimator replaces the single-sequence spectral autocorrelation correction with the multi-chain effective sample size of \citet{Vehtari+etal:2021:Rhat} and propagates the variance to the log scale exactly, which matters in the saturation regime. While the asymptotic properties are theoretically established, Monte Carlo uncertainty is, in our experience, rarely reported, and the MCSE had not previously been validated at scale. Through extensive empirical benchmarking against the \texttt{posteriordb} corpus, we demonstrated that the standard delta-method approximation is highly precise for stable posteriors. Crucially, we identified a structural saturation cap: the MCSE estimate cannot exceed $\approx 1.05$ regardless of how badly the estimator has failed, so values near the cap must be read as saturation rather than as an error bar, and the true error may be far larger. Because calibration already degrades throughout $\operatorname{MCSE} \in (0.3, 1.05)$, the actionable rule is the trust threshold $\operatorname{MCSE}(\log \widehat{Z}) < 0.3$.

Second, we introduced a hybrid score-matching proposal that leverages affine-invariant Riemannian geometry to blend the empirical sample covariance with score-based Fisher information. This regularization approach addresses the operator-norm errors inherent in standard sample covariances, mitigating the severe tail mismatch often encountered in non-Gaussian and moderate-to-high-dimensional posteriors. 

Two limitations delimit the scope of our results. First, our experiments do not cover the rank-deficient regime $S_{\mathrm{fit}} \le p$, and we make no claims about it. The low-rank-plus-diagonal preconditioner of \citet{seyboldt_preconditioning_2026} is a natural extension for that regime. Second, covariance regularization does not address mode coverage, and our corpus contains only one strongly multimodal posterior, the low-dimensional Gaussian mixture. Warp-U--style transformations \citep{wang_warpU_2022} remain the appropriate tool for such targets.

We recommend integrating these tools into a standard evidence-based workflow: default to the hybrid proposal to stabilize the proposal covariance and report the analytic MCSE as a precision metric (targeting $\operatorname{MCSE}(\log \widehat{Z}) < 0.3$).

\section{Acknowledgments}
We thank Ozan Adıg\"uzel for making some of the initial experiments (not shown in this paper), and Noa Kallioinen and Anna Riha for helpful comments and feedback. We also acknowledge the computational resources provided by the Aalto Science-IT project.

\section{Funding}
We acknowledge the support by the Research Council of Finland Flagship programme: Finnish Center for Artificial Intelligence, Research Council of Finland project (340721), and the Aalto Science Institute international summer research internship programme.

\section{Disclosure statement}\label{disclosure-statement}
The authors report there are no competing interests to declare.

\section{Data Availability Statement}\label{data-availability-statement}
The data and code that support the findings of this study are available at the following URL: \url{https://github.com/GiorgioMB/bridgesampling_paper_code}.
The MCSE estimator of Section \ref{sec:mcse} is also implemented in the \texttt{bridgesampling} package \url{https://github.com/quentingronau/bridgesampling/}.

\bibliography{references}

@article{zhang_pathfinder_2022,
	title = {Pathfinder: {Parallel} quasi-{Newton} variational inference},
	volume = {23},
	doi = {10.48550/arXiv.2108.03782},
	language = {English},
	number = {306},
	journal = {Journal of Machine Learning Research},
	author = {Zhang, Lu and Carpenter, Bob and Gelman, Andrew and Vehtari, Aki},
	month = oct,
	year = {2022},
	pages = {1--49},
}

@InProceedings{Magnusson-etal:2025:posteriordb,
  title = 	 {posteriordb: Testing, Benchmarking and Developing {Bayesian} Inference Algorithms},
  author =       {Magnusson, M{\aa}ns and Torgander, Jakob and B{\"u}rkner, Paul-Christian and Zhang, Lu and Carpenter, Bob and Vehtari, Aki},
  booktitle = 	 {Proceedings of The 28th International Conference on Artificial Intelligence and Statistics},
  pages = 	 {1198--1206},
  year = 	 2025,
  editor = 	 {Li, Yingzhen and Mandt, Stephan and Agrawal, Shipra and Khan, Emtiyaz},
  volume = 	 258,
  series = 	 {Proceedings of Machine Learning Research},
  publisher =    {PMLR},
}

@article{gronau_bridgesampling_2020,
	title = {bridgesampling: {An} {R} {Package} for {Estimating} {Normalizing} {Constants}},
	volume = {92},
	doi = {10.18637/jss.v092.i10},
	number = {10},
	journal = {Journal of Statistical Software},
	author = {Gronau, Quentin and Singmann, Henrik and Wagenmakers, Eric-Jan},
	month = feb,
	year = {2020},
}

@Article{gronau_computing_2020,
author={Gronau, Quentin F.
and Heathcote, Andrew
and Matzke, Dora},
title={Computing Bayes factors for evidence-accumulation models using Warp-III bridge sampling},
journal={Behavior Research Methods},
year={2020},
month={Apr},
day={01},
volume={52},
number={2},
pages={918-937},
issn={1554-3528},
doi={10.3758/s13428-019-01290-6},
url={https://doi.org/10.3758/s13428-019-01290-6}
}

@article{koltchinskii_inequalities_2014,
author = {Koltchinskii, Vladimir and Lounici, Karim},
year = {2014},
month = {04},
pages = {},
title = {Concentration Inequalities and Moment Bounds for Sample Covariance Operators},
volume = {23},
journal = {Bernoulli},
doi = {10.3150/15-BEJ730}
}

@misc{seyboldt_preconditioning_2026,
      title={Preconditioning Hamiltonian Monte Carlo by minimizing Fisher Divergence}, 
      author={Adrian Seyboldt and Eliot L. Carlson and Bob Carpenter},
      year={2026},
      eprint={2603.18845},
      archivePrefix={arXiv},
      primaryClass={stat.CO},
      url={https://arxiv.org/abs/2603.18845}, 
}

@article{vehtari_pareto_2024,
	title = {Pareto {Smoothed} {Importance} {Sampling}},
	volume = {25},
	url = {https://arxiv.org/abs/1507.02646v9},
	doi = {10.48550/arXiv.1507.02646},
	number = {72},
	journal = {Journal of Machine Learning Research},
	author = {Vehtari, Aki and Simpson, Daniel and Gelman, Andrew and Yao, Yuling and Gabry, Jonah},
	year = {2024},
	pages = {1--58},
}

@article{bennett_efficient_1976,
	title = {Efficient estimation of free energy differences from {Monte} {Carlo} data},
	volume = {22},
	doi = {10.1016/0021-9991(76)90078-4},
	language = {English},
	number = {2},
	journal = {Journal of Computational Physics},
	author = {Bennett, Charles H},
	month = oct,
	year = {1976},
	pages = {245--268},
}

@article{fruhwirth-schnatter_estimating_2004,
	title = {Estimating marginal likelihoods for mixture and {Markov} switching models using bridge sampling techniques},
	volume = {7},
	url = {https://www.jstor.org/stable/23115004},
	language = {English},
	number = {1},
	journal = {The Econometrics Journal},
	author = {Frühwirth-Schnatter, Sylvia},
	month = feb,
	year = {2004},
	pages = {143--167},
}

@article{gronau_tutorial_2017,
	title = {A tutorial on bridge sampling},
	volume = {81},
	issn = {0022-2496},
	url = {https://www.ncbi.nlm.nih.gov/pmc/articles/PMC5699790/},
	doi = {10.1016/j.jmp.2017.09.005},
	journal = {Journal of Mathematical Psychology},
	author = {Gronau, Quentin F. and Sarafoglou, Alexandra and Matzke, Dora and Ly, Alexander and Boehm, Udo and Marsman, Maarten and Leslie, David S. and Forster, Jonathan J. and Wagenmakers, Eric-Jan and Steingroever, Helen},
	year = {2017},
	pages = {80--97}
}

@article{meng_simulating_1996,
	title = {Simulating {Ratios} of {Normalizing} {Constants} {Via} a {Simple} {Identity}: {A} {Theoretical} {Exploration}},
	volume = {6},
	issn = {1017-0405},
	shorttitle = {Simulating {Ratios} of {Normalizing} {Constants} {Via} a {Simple} {Identity}},
	url = {https://www.jstor.org/stable/24306045},
	number = {4},
	urldate = {2024-01-23},
	journal = {Statistica Sinica},
	author = {Meng, Xiao-Li and Wong, Wing Hung},
	year = {1996},
	note = {Publisher: Institute of Statistical Science, Academia Sinica},
	pages = {831--860},
}

@Misc{rstanarm,
    title = {rstanarm: {Bayesian} applied regression modeling via {Stan}.},
    author = {Ben Goodrich and Jonah Gabry and Imad Ali and Sam Brilleman},
    note = {R package version 2.32.1},
    year = 2024,
    url = {https://mc-stan.org/rstanarm/},
  }

@article{Piironen+Vehtari:2017:rhs,
  title={Sparsity information and regularization in the horseshoe and other shrinkage priors},
  author={Piironen, Juho and Vehtari, Aki},
  journal={Electronic Journal of Statistics},
  volume={11},
  pages={5018--5051},
  year={2017}
}

@article{fourment_dubious_2019,
    author = {Fourment, Mathieu and Magee, Andrew F and Whidden, Chris and Bilge, Arman and Matsen, Frederick A, IV and Minin, Vladimir N},
    title = {19 Dubious Ways to Compute the Marginal Likelihood of a Phylogenetic Tree Topology},
    journal = {Systematic Biology},
    volume = {69},
    number = {2},
    pages = {209-220},
    year = {2019},
    month = {08},
    issn = {1063-5157},
    doi = {10.1093/sysbio/syz046},
    url = {https://doi.org/10.1093/sysbio/syz046},
    eprint = {https://academic.oup.com/sysbio/article-pdf/69/2/209/33861481/syz046.pdf},
}

@Article{wong_splitting_2020,
author={Wong, Jackie S. T.
and Forster, Jonathan J.
and Smith, Peter W. F.},
title={Properties of the bridge sampler with a focus on splitting the {MCMC} sample},
journal={Statistics and Computing},
year={2020},
month={Jul},
day={01},
volume={30},
number={4},
pages={799-816},
issn={1573-1375},
doi={10.1007/s11222-019-09918-5},
url={https://doi.org/10.1007/s11222-019-09918-5}
}

@article{wang_warpU_2022,
author = {Lazhi Wang and David E. Jones and Xiao-Li Meng},
title = {Warp Bridge Sampling: The Next Generation},
journal = {Journal of the American Statistical Association},
volume = {117},
number = {538},
pages = {835--851},
year = {2022},
publisher = {ASA Website},
doi = {10.1080/01621459.2020.1825447},


URL = { 
    
        https://doi.org/10.1080/01621459.2020.1825447
    
    

},
eprint = { 
    
        https://doi.org/10.1080/01621459.2020.1825447
    
    

}
}

@article{meng_schilling_2002,
author = {Xiao-Li Meng and Stephen Schilling},
title = {Warp Bridge Sampling},
journal = {Journal of Computational and Graphical Statistics},
volume = {11},
number = {3},
pages = {552--586},
year = {2002},
publisher = {ASA Website},
doi = {10.1198/106186002457},


URL = { 
    
        https://doi.org/10.1198/106186002457
    
    

},
eprint = { 
    
        https://doi.org/10.1198/106186002457
    
    

}

}

@article{overstall_forster_2010,
title = {Default Bayesian model determination methods for generalised linear mixed models},
journal = {Computational Statistics \& Data Analysis},
volume = {54},
number = {12},
pages = {3269-3288},
year = {2010},
issn = {0167-9473},
doi = {https://doi.org/10.1016/j.csda.2010.03.008},
url = {https://www.sciencedirect.com/science/article/pii/S0167947310001106},
author = {Antony M. Overstall and Jonathan J. Forster}
}

@article{gelman_meng_1998,
author = {Andrew Gelman and Xiao-Li Meng},
title = {{Simulating normalizing constants: from importance sampling to bridge sampling to path sampling}},
volume = {13},
journal = {Statistical Science},
number = {2},
publisher = {Institute of Mathematical Statistics},
pages = {163 -- 185},
year = {1998},
doi = {10.1214/ss/1028905934},
URL = {https://doi.org/10.1214/ss/1028905934}
}

@article{friel_wyse_2012,
author = {Friel, Nial and Wyse, Jason},
title = {Estimating the evidence – a review},
journal = {Statistica Neerlandica},
volume = {66},
number = {3},
pages = {288-308},
doi = {https://doi.org/10.1111/j.1467-9574.2011.00515.x},
url = {https://onlinelibrary.wiley.com/doi/abs/10.1111/j.1467-9574.2011.00515.x},
eprint = {https://onlinelibrary.wiley.com/doi/pdf/10.1111/j.1467-9574.2011.00515.x},
year = {2012}
}

@article{Vehtari+etal:2021:Rhat,
  title={Rank-normalization, folding, and localization: An improved $\widehat{R}$ for assessing convergence of {MCMC}},
  author={Vehtari, Aki and Gelman, Andrew and Simpson, Daniel and Carpenter, Bob and B{\"u}rkner, Paul Christian},
  journal={Bayesian Analysis},
  year=2021,
  volume=16,
 pages={667--718}
}

@article{Beirlant+Bladt:2025,
  author = {Beirlant, Jan and Bladt, Martin},
  title = {Tail classification using non-linear regression on model plots},
  journal = {Extremes},
  volume={28},
  number={2},
  pages={345--369},
  year = {2025}
}

\appendix
\renewcommand{\thetable}{A.\arabic{table}}
\setcounter{table}{0}

\section{\texorpdfstring{\texttt{posteriordb}}{posteriordb} Posteriors and Birthdays Posterior}
\label{app:posteriordb}

We used 41 posteriors from \texttt{posteriordb} and a multi-component Gaussian process model (gpbf6), which was used by \citet{zhang_pathfinder_2022}. Table~\ref{tab:posteriordb} lists the \texttt{posteriordb} IDs, the posterior names used in the figures in this paper, the number of posterior dimensions (in unconstrained space), and a short description of the model.

\begingroup
\setlength{\tabcolsep}{0pt}
\footnotesize
\begin{xltabular}{\textwidth}{@{}p{0.38\textwidth}p{0.2\textwidth}p{0.06\textwidth}p{0.36\textwidth}@{}}
\caption{Posteriors used in Section \ref{sec:results}: 41 from \texttt{posteriordb} plus the Birthdays GP (no \texttt{posteriordb} ID, marked ``-'').
Columns: \texttt{posteriordb} ID, name used in the plots, number of unconstrained posterior dimensions $p$ and a brief description.
`State-votes hier-GP' is listed for completeness but excluded from the calibration analysis (no trustworthy reference; see Section \ref{sec:results}).}%
\label{tab:posteriordb}\\
\toprule
\texttt{posteriordb} id & plot names & $p$ & description \\ \midrule
\endfirsthead
\texttt{posteriordb} id & plot names & $p$ & description \\ \midrule
   \endhead
\bottomrule
\endfoot   
\bottomrule
\endlastfoot   
   arK-arK & AR(5) & 7 & AR-5 model \\
arma-arma11 & ARMA(1,1) & 4 & ARMA \\
bball\_drive\_event\_1-\newline
hmm\_drive\_1 & Bball HMM & 6 & Hidden Markov model \\
bball\_drive\_event\_0-\newline
hmm\_drive\_0 & Bball0 HMM & 6 & Hidden Markov model \\
- & Birthdays GP & 438 & Multi-component GP time series model (gpbf6) \\
butterfly-multi\_occupancy & Butterfly & 106 & Multiple Species-site occupancy model \\
diamonds-diamonds & Diamonds & 26 & Multiple highly correlated predictors log-log model \\
dogs-dogs & Dogs & 3 & Logistic mixed effects model \\
dogs-dogs\_log & Dogs Log & 2 & Logarithmic mixed effects model \\
earnings-logearn\_interaction & Earnings & 5 & Multiple predictors interacting log-linear model \\
eight\_schools-eight\_schools\_\newline
centered & 8-Schools CP & 10 & A centered hierarchical model for 8 schools \\
eight\_schools-eight\_schools\_noncentered & 8-Schools NCP & 10 & A non-centered hierarchical model for 8 schools \\
garch-garch11 & GARCH & 4 & Generalized autoregressive conditional heteroscedastic model \\
GLM\_Binomial\_data-GLM\_Binomial\_model & GLM Binomial & 3 & Success rate of peregrine broods \\
GLM\_Poisson\_Data-GLM\_Poisson\_model & GLM Poisson & 4 & Poisson GLM for modeling a population of peregrines \\
GLMM\_data-GLMM1\_model & GLMM Poisson 1 & 237 & GLMM for peregrine population size \\
GLMM\_Poisson\_data-GLMM\_Poisson\_model & GLMM Poisson 2 & 45 & GLMM for peregrine population size with random site and year effects \\
gp\_pois\_regr-gp\_regr & GP Normal & 3 & Gaussian process regression \\
gp\_pois\_regr-gp\_pois\_regr & GP Poisson & 13 & Gaussian process Poisson regression \\
hmm\_example-hmm\_example & HMM & 4 & Hidden Markov model \\
hudson\_lynx\_hare-lotka\_volterra & Hudson Lynx Hare & 8 & Lotka-Volterra Model \\
irt\_2pl-irt\_2pl & 2PL IRT & 144 & Two-parameter logistic item response theory model \\
fims\_Aus\_Jpn\_irt-2pl\_latent\newline
\_reg\_irt & Japan IRT & 531 & Two-parameter logistic item response theory model \\
low\_dim\_gauss\_mix-low\_dim\_gauss\_mix & Low Dim Gauss Mix & 5 & A Two-dimensional Gaussian mixture model \\
lsat\_data-lsat\_model & LSAT Rasch & 1006 & Random effects (Rasch) model for true difficulty of LSAT questions \\
Mb\_data-Mb\_model & Capture-recapture Mb & 3 & Inferring population size considering immediate trap-response \\
mcycle\_gp-accel\_gp & Mcycle GP & 66 & Heteroscedastic Gaussian processes \\
Mh\_data-Mh\_model & Capture-recapture Mh & 388 & Logistic-normal heterogeneity model \\
Mtbh\_data-Mtbh\_model & Capture-recapture Mtbh & 154 & Capture-recapture model where detection probability varies by occasion \\
Mth\_data-Mth\_model & Capture-recapture Mth & 394 & Capture-recapture model where detection probability varies by occasion and heterogeneity \\
nes2000-nes & NES2000 & 10 & Multiple predictor linear model \\
one\_comp\_mm\_elim\_abs-one\_comp\_mm\_elim\_abs & PKPD One Comp & 4 & One compartment PKPD model \\
radon\_all-\newline
radon\_hierarchical\_intercept\newline
\_noncentered & Radon Hierarchical & 391 & A county intercept and county level covariate for Radon data (non-centered) \\
radon\_all-\newline 
radon\_variable\_intercept\_\newline
\_slope\_noncentered & Radon Intercept & 777 & Variable intercept and slope hierarchical for Radon data (noncentered) \\
radon\_all-\newline
radon\_partially\_pooled\_\newline
\_noncentered & Radon Pooled & 389 & Hierarchical intercept model for Radon data (non-centered) \\
radon\_all-\newline 
radon\_variable\_slope\_\newline
\_noncentered & Radon Slope & 390 & Variable slope hierarchical for Radon data (non-centered) \\
sat-hier\_2pl & SAT Rasch & 670 & Hierarchical two-parameter logistic item response model \\
sblrc-blr & sblrc & 6 & A Bayesian linear regression model with vague priors \\
science\_irt-grsm\_latent\_reg\_irt & Science IRT & 408 & Rating scale and generalized rating scale models with latent regression \\
state\_wide\_presidential\_votes-hierarchical\_gp & State-votes hier-GP & 933 & Hierarchical Gaussian process \\
Survey\_data-Survey\_model & Survey & 1 & Inferring the return rate and number of surveys from observed returns \\
timssAusTwn\_irt-gpcm\_latent\newline 
\_reg\_irt & timssAusTwn IRT & 530 & Partial credit and generalized partial credit models with latent regression \\
 \end{xltabular}
\endgroup

\end{document}